\documentclass{article}

\begin{document}

\title{Closed Trapped Surfaces in Cosmology}
\author{George F\ R\ Ellis \\
 Mathematics Department, University of Cape Town}

\maketitle

\begin{abstract}
The existence of closed trapped surfaces need not imply a cosmological
singularity when the spatial hypersurfaces are compact. This is illustrated
by a variety of examples, in particular de Sitter spacetime admits many
closed trapped surfaces and obeys the null convergence condition but is
non-singular in the k=+1 frame.
\end{abstract}

\section{Introduction}

Since Roger Penrose' pioneering paper of 1965 \cite{pen65}, the existence of
closed trapped surfaces (`CTSs') has been understood as a geometrical
condition\ that, jointly with suitable energy conditions, in many
circumstances leads to the existence of space-time singularities in the
context of both gravitational collapse and cosmology. This understanding has
been codified in the series of singularity theorems proved by Penrose and
Stephen Hawking \cite{pen65,hawpen70,hawell73}. The various theorems involve
different combinations of geometric requirements and energy conditions.
Under the assumptions of standard hot big bang theory, these conditions will
indeed be met in the cosmological context, because the existence of the
black-body cosmic background radiation implies the existence of CTSs in the
era between decoupling and the present day \cite{hawell68,hawpen70,hawell73}
and so leads to prediction of a (classical) singularity at the start of the
universe. However it is now known that scalar fields can violate some of the
energy conditions, thereby providing the foundation of the inflationary
universe paradigm \cite{inflate}. The possibility then arises of singularity
avoidance in realistic early universe models because of these energy
condition violations, and indeed even avoidance of a quantum gravity regime
is possible \cite{ell03}, despite the existence of CTSs. The purpose of this
paper is to revisit the relation between CTSs and spacetime singularities in
this cosmological context, characterising cases where existence of CTSs do
not imply the existence of singularities. In particular we examine the case
of the de Sitter Universe, showing that CTSs exist in these spacetimes even
though (in maximally extended form) they are both geodesically complete and
stable to perturbations.

\subsection{ Closed Trapped Surfaces}

Consider a spacelike 2-surface S with spherical topology. It is a \textit{\
past closed trapped surface} if\ both families of \ past null geodesics
orthogonal to the 2-surface S are converging, i.e. if their divergences
(evaluated at the surface) are negative. Similarly \textit{future closed
trapped surfaces} occur if we replace ''past''\ by ''future''\ in the above.
Both past and future closed trapped surfaces will be referred to as \textit{
closed trapped surfaces} (`CTSs'). We will in this paper be concerned with
2-surfaces that are 2-spheres with a group of isometries transitive on them,
so they are homogeneous 2-dimensional subspaces of spacetime. Then the value
of the divergence of each family of normals is constant over the 2-surface,
and can be characterised by a single number on each 2-sphere. Thus we will
in fact be considering existence of homogeneous closed trapped 2-spheres.

\textit{Marginally closed trapped surfaces} exist if the divergences are
non-positive for these families of null geodesics; that is, if the
divergences are either zero or negative rather than strictly negative.

\subsection{Energy Conditions}

Energy conditions conditions generically lead to convergences of
irrotational familes of non-spacelike and null geodesics respectively.

\subsubsection{Non-spacelike convergence condition}

This is the condition

\[
R_{ab}K^{a}K^{b}\geq 0\mathrm{~ for~ all~ non-spacelike~ vectors~ }K^{a}. 
\]
For perfect fluids, this translates into $\mu +p\geq 0,$ $\mu +3p\geq 0,$
which will be true for all ordinary matter. For scalar fields, it becomes $
\frac{1}{2}\dot{\phi}^{2}\geq 0,\ \dot{\phi}^{2}\geq V(\phi ),$ hence is
violated when the slow rolling condition $\ \dot{\phi}^{2}\ll V(\phi )$ is
satisfied. A cosmological constant is the case $\ \dot{\phi}^{2}=0,\ V>0$
and hence violates this condition.

\subsubsection{Null convergence condition}

This is the condition

\begin{equation}
R_{ab}K^aK^b\geq 0\mathrm{~for ~all ~null ~vectors~ }K^a  \label{ncc}
\end{equation}
which is implied by the previous:\ 

\[
\{\mathrm{{Non-spacelike~ convergence~ condition~}\}\Rightarrow \{{~Null
~convergence~ condition}\}. } 
\]
For perfect fluids, this translates into $\mu +p\geq 0,$ which will be true
for all ordinary matter. For scalar fields, it is $\frac{1}{2}\dot{\phi}
^{2}\geq 0,$ and so is true for all ordinary scalar fields (we discount the
possibility of `phantom matter' that violates this condition, see Gibbons 
\cite{gib03} for a discussion).

\subsection{Focussing}

The equation determining the evolution of the convergence $\theta
=K_{;a}^{a} $ of hypersurface-orthogonal null geodesics is

\begin{equation}
\frac{d\theta }{dv}+\frac{1}{2}\theta ^{2}=-R_{ab}K^{a}K^{b}-2\sigma ^{2},
\label{nullray}
\end{equation}
while the shear propagation equation is

\[
\frac{d}{dv}\sigma _{mn}=-\theta \sigma _{mn}-C_{manb}K^{a}K^{b}. 
\]
This shows that the shear can only remain zero either for very special
spacetimes (e.g. Robertson-Walker spacetimes where $C_{manb}=0)$, or for
very special null rays in a more generic spacetime, so that $
C_{manb}K^{a}K^{b}=0$ at every point on the null geodesics because the
geodesic tangent vector is in a special relation to the Weyl tensor (it is a
principal null direction). Once the shear is non-zero, it acts as a source
term in the null Raychaudhuri equation (\ref{nullray}).

When the null convergence condition (\ref{ncc}) is true, there is an
exceptional case and a generic case for familes of hypersurface orthogonal
null geodesics. The exceptional case occurs if $\theta =0,\,$i.e. no
focussing occurs: 
\[
\theta =0\Rightarrow R_{ab}K^aK^b=0,\ \sigma ^2=0\Rightarrow
C_{acbd}K^aK^b=0. 
\]
This cannot be true in a generic cosmological context, for example a
perturbed Robertson-Walker universe, when both the Ricci and Weyl tensor
conditions will be violated along a generic null ray. The generic case is
when either $\theta _0<0$ or $\theta _0>0$. Both imply $\theta \rightarrow
-\infty $ \ within a finite affine distance, either to the past or the
future. Then the null rays intersect at a caustic, so the surface generated
by the null geodesics experiences self-intersections either before or at
those events.

Hence when we consider CTSs in realistic cosmologies, which will always
satisfy the null convergence condition, both families of converging
orthogonal null geodesics, which generate the boundary of their past for the
case of past CTSs, will self-intersect. Then by well-known causal theorems,
these null rays will lie from then on inside the pasts of the CTSs; the
boundary of the past is therefore compact. This is what underlies the
singularity theorems.

\subsection{The Major Singularity Theorems}

The major singularity theorems referring to CTSs are given in Hawking and
Ellis \cite{hawell73} (`HE'). Each case assumes a CTS\ or roughly equivalent
condition, plus the following:

\textbf{HE Theorem 1} \cite{pen65} - A non-compact Cauchy surface and the
null convergence condition,

\textbf{HE\ Theorem 2} \cite{hawpen70} - The non-spacelike convergence
condition and a causality condition,

\textbf{HE\ Theorem 3 }\cite{haw67} - The non-spacelike convergence
condition and a causality condition.

Roughly speaking: in each case the boundary of the past comes to an end
because of existence of self-intersections points in its generating
geodesics; but the past is contained within the boundary, hence a
singularity must occur.

We now consider how these theorems apply to Friedmann-Lemaitre (`FL')\ model
universes, interpreted here as universes with a Robertson-Walker geometry
and matter content of ordinary matter and/or a scalar field.

\section{Friedmann-Lem\^{a}itre Models}

Friedmann-Lemaitre universe have a Robertson-Walker (`RW') metric which can
be represented in the form 
\begin{equation}
ds^2=-dt^2+S^2(t)\left( dr^2+f^2(r)(d\theta ^2+\sin ^2\theta d\phi ^2)\right)
\label{rw}
\end{equation}
The scale factor is $S(t)$, the matter 4-velocity is: $u^a=\delta
_0^a\Rightarrow u_a=g_{ab}u^b=-\delta _a^0,$ and $f(r)=(\sin r,r,\sinh r)$
if ($k=1,0,-1)\;$respectively$.$

The metric determinant $g$ is 
\[
g=-S^{6}(t)f^{4}(r)\sin ^{2}\theta \Rightarrow \sqrt{-g}=S^{3}(t)f^{2}(r)
\sin \theta . 
\]
Note that every 2-surface $\mathcal{S}(t,r):(t=const,r=const)$ is a
homogeneous 2-sphere of area $\ \ $

\[
A=4\pi S^{2}(t)f^{2}(r). 
\]

\subsection{Radial Null Geodesics}

The family of past-directed radial null geodesics in this space-time has
tangent vector field

\[
K^a=\frac 1{S(t)}\left( -1,\pm \frac 1{S(t)},0,0\right) =\frac{dx^a}{dv}
\]
\cite{ell71} where the sign depends on whether the geodesics are ingoing or
outgoing. This form gives 
\[
K^aK_a=g_{ab}K^aK^b=0,
\]
\[
K^au_a=\frac 1{S(t)}=(1+z)
\]
as required, and they are normal to the family of instantaneous homogeneous
2-spheres $\mathcal{S}(t,r)$ because they lie in the orthogonal 2-plane $
(\theta =const,\phi =const)$ to these 2-spheres, described by coordinates $
(t,r).$ They diverge from the point of origin of coordinates $r=0,$ and
refocus at the antipodal point $r=\pi $ when $k=+1.$

The divergence of this family of null geodesics is given by

\begin{equation}
K_{;a}^a=\frac 1{\sqrt{-g}}\frac \partial {\partial x^a}(\sqrt{-g}K^a)=\frac
2{S^2(t)}\left[ -\dot{S}(t)\pm \frac{\partial f(r)/\partial r}{f(r)}\right] .
\label{div}
\end{equation}

CTSs occur if the divergence is negative for both families of null geodesics
for some value of $r$ and $t$, i.e. for both choices of sign in the last
term on this 2-surface.We now look at a series of specific cases.

\section{The Early and Late Universe}

\subsection{Standard radiation/matter dominated expansion}

Here $S(t)=at^n$ with $a>0$ and $n=2/3$ in the matter era, $n=1/2$ in the
radiation era, and with $0<t<\infty $. The divergence of the radial families
of null geodesics is given by (\ref{div}): 
\[
K_{;a}^a=\frac 2{at^{2n}}\left[ -ant^{n-1}\pm (\cot r,\frac 1r,\coth
r)\right] 
\]
for $k=+1,0.-1$ respectively. These expressions give the divergences of the
normals of the 2-spheres $(r,t)$ constant. For each $t>0$ and each value of $
k$ these will be CTSs, obtained by choosing $r$ large enough that the
magnitude of the second term in the square brackets is less than that of the
first term in these brackets, so that $K_{;a}^a<0$ for both signs (i.e. for
both ingoing and outgoing null geodesics). Note that previous examinations
of these surfaces have concentrated on showing that the past light cone of
any observer will start reconverging and thus that closed trapped surfaces
will occur in the past of the observer, associated with this refocussing of
their past light cone. What is shown here is a bit different, namely that
for any value of $t,$ if one goes to large enough values of $r$ there will
be closed trapped surfaces surrounding the origin at that time. Hence there
are closed trapped surfaces surrounding us even today (associated with the
reconvergence of the past light cone of observers in our future).

Here the energy condition $\mu +3p>0$ is satisfied because we consider only
ordinary matter. If this energy condition does not change in the past then a
singularity is predicted in all cases via the Raychaudhuri equation for
exact FL\ models \cite{ell03}, and via Theorems 2\&3 above for universes
that are perturbed FRW\ models at late times even if they are quie different
at early times (small enough perturbations will necessarily preserve the
inequalities $K_{;a}^a<0$ in the late-time era), and also by Theorem 1 in
the case $k\neq 1$. Hence the existence of singularities is predicted and is
stable to perturbations of these models at late times (which may correspond
to very large changes in the models at early enough times) \cite{hawell73}.

\subsection{Inflation}

In realistic universes the Hot Big Bang\ era may be preceded by an era of
inflation. It will still be true that CTSs\ occur in the late universe
(between decoupling and the present day, as well as in the Hot Big Bang era
itself). However inflation violates the timelike convergence condition at
early times and so can avoid the initial singularity that would otherwise be
predicted because these CTSs exist at later times. Theorems 2 and 3 fall
away because of the timelike energy condition violation. If $k=0$ or $-1$,
then a singularity will indeed occur, because of the Friedmann equation in
the exact FL\ case \cite{ell03}, and in the perturbed FL\ case because of
Theorem 1, relying only on the null energy condition, together with the
existence of open space sections if they have their normal topology.
Singularities are not inevitable when $k=+1,$ both because the Friedmann
equation now allows a minimum and because there are then closed spacelike
sections so none of the above theorems apply. Various kinds of non-singular
model can then occur \cite{ellmaa03}

\section{The de Sitter Universe}

We deal in turn with the three RW frames for de Sitter spacetime (see \cite
{rob33,sch56,hawell73} for its global properties).

\subsection{The k=+1 Frame}

In the global $k=+1$\ frame, the metric is (\ref{rw}) with $S(t)=A\cosh Ht$, 
$\ f(r)=\sin r,$ where $A,$ $H>0$ and $-\infty <t<\infty .$ The cosmology is
non-singular and geodesically complete. Note that $0\leq r\leq \pi .$ The
antipodal point to the origin of coordinates is at $r=\pi ;$the equator for
these coordinates is at $r=\pi /2.$ Then by  (\ref{div})

\[
K_{;a}^a=\frac 2{A^2\cosh ^2Ht}\left[ -AH\sinh Ht\pm \frac{\cos r}{\sin r}
\right] 
\]
We get a past closed trapped surface if $t>0$ and $\ AH\sinh Ht>|\frac{\cos r
}{\sin r}|.$ Now the latter term is zero at $r=\pi /2$ and diverges
positively and negatively respectively at $r=0,\pi .$ Thus provided $t>0,$
we have a past closed trapped surface for all $r$ such that 
\[
AH\sinh Ht>|\cot r|
\]
which defines a non-zero set of 2-spheres around the equator at $r=\pi /2$.
As $t\rightarrow 0,$ these shrink to just the equator; as $t\rightarrow
\infty $ , they expand to a large part of the whole sphere. The origin is
arbitrary, so every 2-sphere $(t,r)$ constant with $t>0$ and area greater
than 
\[
A_{*}=4\pi A^2\cosh ^2Ht\sin ^2(r_{*}),~~AH\sinh Ht=\cot r_{*}
\]
in any $k=+1$ frame will be a past closed trapped surface. Its normals will
self-intersect and have caustics where $K_{;a}^a\rightarrow \infty $ at both 
$r\rightarrow 0$ (the origin, the ingoing family) and $r\rightarrow \pi $
(the antipodal point, the outgoing family)$.$ Since $\sin ^2r=\frac 1{1+\cot
^2r}$ this gives 
\[
A_{*}=4\pi \frac{A^2\cosh ^2Ht}{1+A^2H^2\sinh ^2Ht}.
\]
As $t\rightarrow 0$, $S(t)\rightarrow A^2$ and $A_{*}\rightarrow 4\pi A^2$
(corresponding to $r\rightarrow \pi /2$); as $t\rightarrow \infty $, $
S(t)\rightarrow \infty $ and $A_{*}\rightarrow 4\pi \frac 1{H^2}$, so there
is a minimum radius that will give a CTS (but the size of the 3-spaces
increases without limit, so this minimum radius will be an ever smaller
fraction of the size of the space sections). For $t<0$ we find the
corresponding family of future trapped surfaces. There are no trapped
surface for $t=0.$

For these trapped surfaces (\ref{nullray})\ becomes 
\begin{eqnarray*}
\frac{d\theta }{dv} &=&-\frac 12\theta ^2\Rightarrow \frac{d\theta }{\theta
^2}=-\frac 12dv\Rightarrow \frac 1\theta =-\frac 12(v-v_0) \\
&\Rightarrow &\ ~\theta =-\frac 2v\mathrm{~on~choosing~}v_0=0.
\end{eqnarray*}
Thus the geodesics generating the pasts of the set of past trapped surfaces
locally self-intersect, hence signalling an end to the boundary of the past
of the trapped surfaces. But these intersections occur round the back (near
the antipodal points); hence the past of these surfaces is not trapped by
these null geodesics. Theorem 1 does not apply, even though the null energy
condition is true, because the spatial sections are compact.

Note that these space sections are not unique \cite{sch56}: every 3-sphere
passing through the de Sitter throat is equivalent to every other one. The
same result as above must be found in every such frame. This at first seems
to lead to an apparent contradiction:\ no consistency is found at the same
points on the hyperboloid in different frames, because in the $t=0$ frames
of different such choices there are no trapped surfaces but at the same
points in the $t\neq 0$ frames there do exist closed trapped surfaces.
However \textit{the correspondence between events on the hyperboloid and
2-spheres depends on the frame chosen. }The frame corresponds point by point
to events on the hyperboloid in each coordinate frame in a way such that the
corresponding 2-spheres (spherically surrounding the origin of coordinates)
depends on the coordinate frame chosen. \textit{Given a specific choice of
frame, however, a unique such correspondence of points and 2-spheres exists}
. Those found in the $t=0_{}$surfaces in a particular frame may be
represented as marginally trapped 2-spheres in that frame, but are fully
trapped in other frames. Thus one must choose a specific frame and work it
all out in that frame; the results in all other frames will then follow by
boosting, rotating and translating that frame.

\subsection{Perturbed de Sitter Universes}

The further basic point is that existence of CTSs do not imply singularities
in perturbed de Sitter universes either, when the null generic condition
holds, even when $\mu +p\geq 0$ (so that self-intersections occur because of
the generic conditionds as outlined above) but provided still $\mu +3p<0$.
This is because they don't imply them in the de Sitter case, where the past
of each CTS is also compact ,but that does not imply a singularity because
of the closed space sections. If this were not true the de Sitter universe
would be unstable - but it is well known to be stable.

\subsection{The k=0 frame}

In the $k=0$\ frame, which covers half the spacetime (and so is not
geodesically complete), we have the metric (\ref{rw}) with $S(t)=A\exp Ht$ , 
$\ \ f(r)=r,$ for $-\infty <t<\infty $ and $A,H>0.$ In this case  (\ref{div}
) shows

\[
K_{;a}^a=\frac 2{A^2\exp 2Ht}\left[ -AH\exp Ht\pm \frac 1r\right] 
\]
The second term dies away to zero, so there will be closed trapped surfaces
for 
\[
AH\exp Ht>\frac 1r
\]
which will always be true for large enough $r$ for any $t.$ These will be
part of the same set of 2-spheres as characterised above, but expressed in
different coordinates. In this case they do correspond to geodesic
incompleteness, because this coordinate frame does not cover the whole
hyperboloid and Theorem 1 above applies in this frame. New information could
come in from the other half of the hyperboloid if the solution is extended
further (which even though it lies beyond the infinite redshift surface need
not correspond to infinite redshift of matter beyond that surface; that
would depend on how matter is moving in this further part of the
hyperboloid). The spacetime is singular but extendible.

\subsection{The k=-1 frame}

In the $k=-1$\ frame, which covers less than half the spacetime (and so is
not geodesically complete), we have the metric (\ref{rw})$\ $with $
S(t)=A\sinh Ht$, $\ f(r)=\sinh r,$ for $0<t<\infty $ and $A,H>0.$ In this
case  (\ref{div}) shows that
\[
K_{;a}^a=\frac 2{A^2\sinh ^2Ht}\left[ -AH\cosh Ht\pm \frac{\cosh r}{\sinh r}
\right] 
\]
The latter term is always of magnitude $>1$, diverging as $r\rightarrow 0$
and $\rightarrow 1$ as $r\rightarrow \infty .$ Thus there will be closed
trapped surface for values of time $t$ such that 
\[
AH\cosh Ht>1,\;t>0\Leftrightarrow t>(1/H)\arg \cosh (1/AH)
\]
which will exist for all $A,H>0.$ For those values of $t$, closed trapped
surfaces exist for all values of $r$ such that

\[
AH\cosh Ht>\coth r 
\]
which will then exist for large enough $r.$ These will again be part of the
same set of 2-spheres as characterised above, but expressed in different
coordinates. Here again Theorem 1 predicts an in initial singularity and
again they are geodesically incomplete but extendible.\newline

\subsubsection{Conclusion: de Sitter spacetime}

In all three cases, we show that \textit{past closed trapped surfaces exist
in the de Sitter universe} (of course these are just the same set of
2-surfaces found in different coordinates) and lead to \textit{the
boundaries of the pasts of those 2-surfaces being compact}. This does not
however lead to spacetime singularities in the first case (de Sitter
spacetime is geodesically complete in the $k=+1$ frame), because the pasts
of the trapped 2-spheres are not trapped by these null boundaries, rather
they can escape freely to earlier times because the 3-spaces are compact
(the null cone intersections take place at the antipodal point on the other
side of the 3-spaces with 3-sphere topology, allowing any interior matter to
escape on this side, near the origin). However in the $k=0$ and $k=-1$
frames the cosmology is singular because the worldlines do not cover the
whole spacetime; indeed they self-intersect at $t=0$ in the $k=-1$ case.

\section{Emergent Universes}

These are \ non-singular models with $k=+1$ that start off asymptotically as
Einstein Static universes, and then evolve to de Sitter universes (and at
even later times to a shandard hot big bang) \cite{ellmaa03}. Here $k=+1$
and in the simplest case $S(t)=A+B\exp Ht$, with $-\infty <t<\infty ,$ and $
A,B>0$ \cite{ellmur03}. Then by  (\ref{div}) 
\[
K_{;a}^a=\frac 2{(A+B\exp Ht)^2}\left[ -BH\exp Ht\pm \frac{\cos r}{\sin r}
\right] 
\]
We get a past closed trapped surface if $\ BH\exp Ht>\>|\frac{\cos r}{\sin r}
|.$ Now the latter is zero at $r=\pi /2$ and diverges to -$\infty ,\infty $
at $r=0,\pi \,$respectively$.$Thus provided $t>0,$ we have a closed trapped
surface for all $r$ such that 
\[
BH\exp Ht>|\cot r|
\]
which defines a non-zero set of 2-spheres around the equator, depending on
the coordinate time $t$. As $t\rightarrow -\infty ,$ the static limit, these
shrink to just the equator; as $t\rightarrow \infty $, they expand to a
large part of the whole sphere (as in the de Sitter case). However the
spacetime is geodesically complete and non-singular. The past generators of
the 2-spheres intersect , and this does not imply the exstence of
singularities; none of the singularity theorems apply. The de Sitter
universe is the special case when $A=0,$ obtained as a completely smooth
limit.

\subsection{Perturbed Emergent universes}

The existence of CTSs does not imply singularities in perturbed emergent
universe eithers. Perturbations that ensure the genericity condition on all
null geodesics again do not imply a singularity, as in the case of the de
Sitter universe.

\section{The Einstein Static Universe}

Here we have a RW metric (\ref{rw})$\;$with $S(t)=S_0>0,$ $k=+1,$ $-\infty
<t<\infty $, and so

\[
K_{;a}^a=\frac 2{S^2(t)}\left[ -\dot{S}(t)\pm \frac{\partial f/\partial r}{
f(r)}\right] =\frac 2{S_0^2\ }\left[ 0\pm \frac{\cos r}{\sin r}\right] .
\]
In this case there are no closed trapped surfaces; however marginally
trapped surfaces occur on the equator, the null geodesics generating their
pasts intersecting at the antipodal pointRoughly:\ the closure of the space
sections does not allow existence of 2-spheres that are large enough to be
trapped. . This lack of closed trapped surfaces is connected to the high
degree of stability of the E-S universe (they are stable to all inhomogeneus
perturbations in the radiation case \cite{esperturb}). Again it is thus true
that perturbing the universe leads to the generic condition for null
geodesics but not to singularities.

\section{Conclusion}

Closed trapped surfaces occur in most Friedmann models, including the de
Sitter universe. They necessarily lead to a singularity only if $\rho +3p>0$
. When $\rho +3p<0$ and $k=+1$, singularity avoidance is possible. The null
energy condition $\rho +p>0$ does not necessarily lead to a singularity,
despite existence of these closed trapped surfaces and hence compact
boundaries of the past of these 2-spheres, when the spatial sections are
compact.


\begin{thebibliography}{99}
\bibitem{pen65} {\small R Penrose (1965), ``Gravitational collapse and
spacetime singularities'', Phys. Rev. Lett. {\bf 14}, 57.}

\bibitem{hawpen70}{\small S W Hawking and R Penrose (1970),
``The singularities of gravitational collapse and cosmology'',
Proc. Roy. Soc. Lond. A {\bf 314}, 529.}

\bibitem{hawell73}{\small S W Hawking and G F R\ Ellis (1973), {\it The
Large Scale Structure of Space-Time} (Cambridge: Cambridge
University Press).}

\bibitem{hawell68}{\small S W Hawking and G\ F\ R Ellis (1968),
`The cosmic black body radiation and the existence of singularities
in our universe'', Astrophys. J. {\bf 152}, 25.}

\bibitem{inflate} {\small A H Guth (1981), ``Inflationary
universe: a possible solution to the horizon and flatness problems'',
Phys. Rev. D {\bf 23}, 347}; {\small A H Guth (2001), ``Eternal inflation'',
astro-ph/0101507.}

\bibitem{ellmaa03}  {\small G F R Ellis, R  Maartens (2002) "Eternal inflation without quantum gravity" gr-qc/0211082 .}

\bibitem{gib03}{\small  G\ W\ Gibbons ''Phantom matter and the cosmological
constant'' hep-th/0302199}

\bibitem{haw67} {\small S W Hawking (1967), ``The occurrence
of singularities in cosmology II: causality and singularities'',
Proc. Roy. Soc. Lond. A {\bf 300}, 187.}

\bibitem{ell71} {\small G F R Ellis (1971), ``Relativistic cosmology'',
in {\em General Relativity and Cosmology\/} ({\em Proc. 47th Enrico
Fermi Summer School\/}), ed R K Sachs, (New York: Academic Press),
104.}

\bibitem{ell03} {\small G F R Ellis (2003), "Cosmological Singularities and Perturbations". Paper presented at Stephen Hawking's 60th Birthday meeting, Cambridge. To appear, proceedings.}

\bibitem{rob33} {\small H P Robertson (1933),
``Foundations of relativistic cosmology'',
Proc. Nat. Acad. Sci. U.S. {\bf 15}, 822.}

\bibitem{sch56} {\small E Schr\"{o}dinger (1956), {\it Expanding
Universes}, (Cambridge: Cambridge University Press).}

\bibitem{esperturb} {\small  J  D Barrow, G F R Ellis, R  Maartens, C  Tsagas (2003) ``On the Stability of the Einstein Static Universe''. gr-qc/0302094 }

\bibitem{ellmur03}  {\small G F R Ellis and J Murugan (2003), ``The Emergent Universe: an explicit Integration''. Preprint.}

\end{thebibliography}
\end{document}